\documentclass[aps,prb,reprint,twocolumn,showpacs,floatfix,superscriptaddress,nofootinbib]{revtex4}
\usepackage{graphicx}
\usepackage{amssymb}
\usepackage{amsmath}
\usepackage{lmodern}
\usepackage{color}
\usepackage{hyperref}
\usepackage{empheq}
\usepackage{epstopdf}
\usepackage{amsmath}

\DeclareMathOperator{\sech}{sech}

\begin{document}

\title{The kinks, the  solitons, the breathers and the shocks in series connected discrete Josephson transmission lines}

\author{Eugene Kogan}
\email{Eugene.Kogan@biu.ac.il}
\affiliation{Department of Physics, Bar-Ilan University, Ramat-Gan 52900, Israel}
\affiliation{Max-Planck-Institut fur Physik komplexer Systeme
Dresden 01187, Germany}
\affiliation{Donostia International Physics Center (DIPC),
Paseo de Manuel Lardizabal 4, 20018 San Sebastian/Donostia, Spain}

\begin{abstract}
We  analytically study the localized running waves  in the discrete Josephson transmission lines (JTL), constructed from  Josephson junctions (JJ) and  capacitors.
The quasi-continuum approximation reduces  calculation of the running wave properties to  the problem of equilibrium of an elastic rod in the potential field.
Making additional approximation, we reduce the problem to
the motion of the fictitious Newtonian particle in the potential well.
We show that there exist running waves in the form of supersonic kinks and    solitons
and
calculate their velocities and  profiles. We show that the nonstationary smooth waves which are small perturbations on the homogeneous non-zero background are described by Korteweg-de Vries equation, and those on  zero background --  by modified Korteweg-de Vries equation.
We also study the effect of dissipation on the running waves in JTL and find that in the presence of the resistors, shunting the JJ and/or in series with the ground capacitors, the only possible stationary running waves are the shock waves, whose  profiles are also found. Finally in the framework of Stocks expansion we study the nonlinear dispersion and modulation stability in the discrete JTL.
\end{abstract}

\date{\today}

\maketitle

\section{Introduction}
\label{introd}

The concept that in a nonlinear wave propagation system
the various parts of the wave travel with different
velocities, and that wave fronts (or tails) can sharpen
into shock waves, is deeply imbedded in the classical
theory of fluid dynamics \cite{whitham}.
The methods developed in that field can be profitably used
to study signal propagation in nonlinear transmission lines
\cite{french,nouri,neto,nikoo,silva,wang,rangel,kyuregyan,akem,fairbanks}.
In the early studies of shock waves in  transmission lines, the
origin of the nonlinearity was due to nonlinear capacitance
in the circuit \cite{landauer,peng,rabinovich}.

Interesting and potentially important examples of nonlinear transmission lines are circuits containing Josephson junctions (JJ) \cite{josephson} -
Josephson transmission lines (JTL) \cite{barone,pedersen,tinkham,kadin}.
The unique nonlinear properties of JTL allow to construct
soliton propagators,
microwave oscillators, mixers, detectors,
parametric amplifiers, and  analog amplifiers \cite{pedersen,kadin,tinkham}.

Transmission lines formed by JJ  connected in series were
studied beginning from  1990s, though much less than transmission lines
formed by JJ  connected in parallel \cite{solitons}.
However, the former began to attract quite a lot of attention  recently \cite{yaakobi,brien,macklin,kochetov,zorin,basko,dixon,goldstein}, especially
in connection with possible JTL traveling wave parametric amplification
\cite{white,miano,pekker}.

The interest in studies of discrete nonlinear electrical transmission lines, in particular of lossy nonlinear transmission lines, has started some time ago \cite{rosenau,chen,mohebbi}, but it became even more pronounced recently
\cite{ricketts,houwe,katayama,sekulic}.
These studies should be seen in the general context of waves in strongly nonlinear
discrete systems \cite{kevrikidis0,english,kevrikidis,nesterenko0,malomed2,nesterenko,malomed}.

In our previous publication \cite{kogan} we  considered  shock waves in the continuous JTL with resistors, studying the influence of those on the  shock profile.
Now we want to analyse wave propagation in the discrete  JTL, both lossless and lossy

The rest of the paper is constructed as follows.
 In  Section~\ref{quasi} we formulate the approximation to the circuite
equations of the  discrete
lossless JTL. In Section~\ref{general} we formulate the quasi-continuum approximation and show the analogy between the problem of the running waves and the problem of equilibrium of an elastic rod in the potential field.  In Section~\ref{shock}, by simplifying the  approximation, we reduce the problem of the running waves  to an effective mechanical problem,  describing motion of a fictitious particle in a potential well and study  the profiles of the kinks and of the   solitons. In Section~\ref{weak}
we consider specifically weak kinks and weak solitons.
In  Section~\ref{shunti} we  discuss the  effect of dissipation  on the  running waves in the discrete JTL.
In Section~\ref{contr} we formulate the modified quasi-continuum approximation and, on top of it, the simple wave approximation, which opens the way to conveniently study non-stationary waves in the JTL.
In Section~\ref{stocks} we obtain the nonlinear dispersion law, and
in Section~\ref{modul} we study the  modulation stability
of the wavetrains.
In Section~\ref{discussion} we briefly mention possible   applications of the results obtained in the paper and opportunities for their generalization.
In the Appendix~\ref{linear} we  apply the modified quasi-continuum approximation to the discrete linear  transmission line.
In the Appendix~\ref{int} we  propose the integral approximation to the discrete  transmission lines equations. In the Appendix~\ref{ape}, added after the paper was published, we show that the approach of the paper allows us  to describe also the breathers.

\section{The discrete Josephson transmission line}
\label{quasi}

\begin{figure}[h]
\includegraphics[width=\columnwidth]{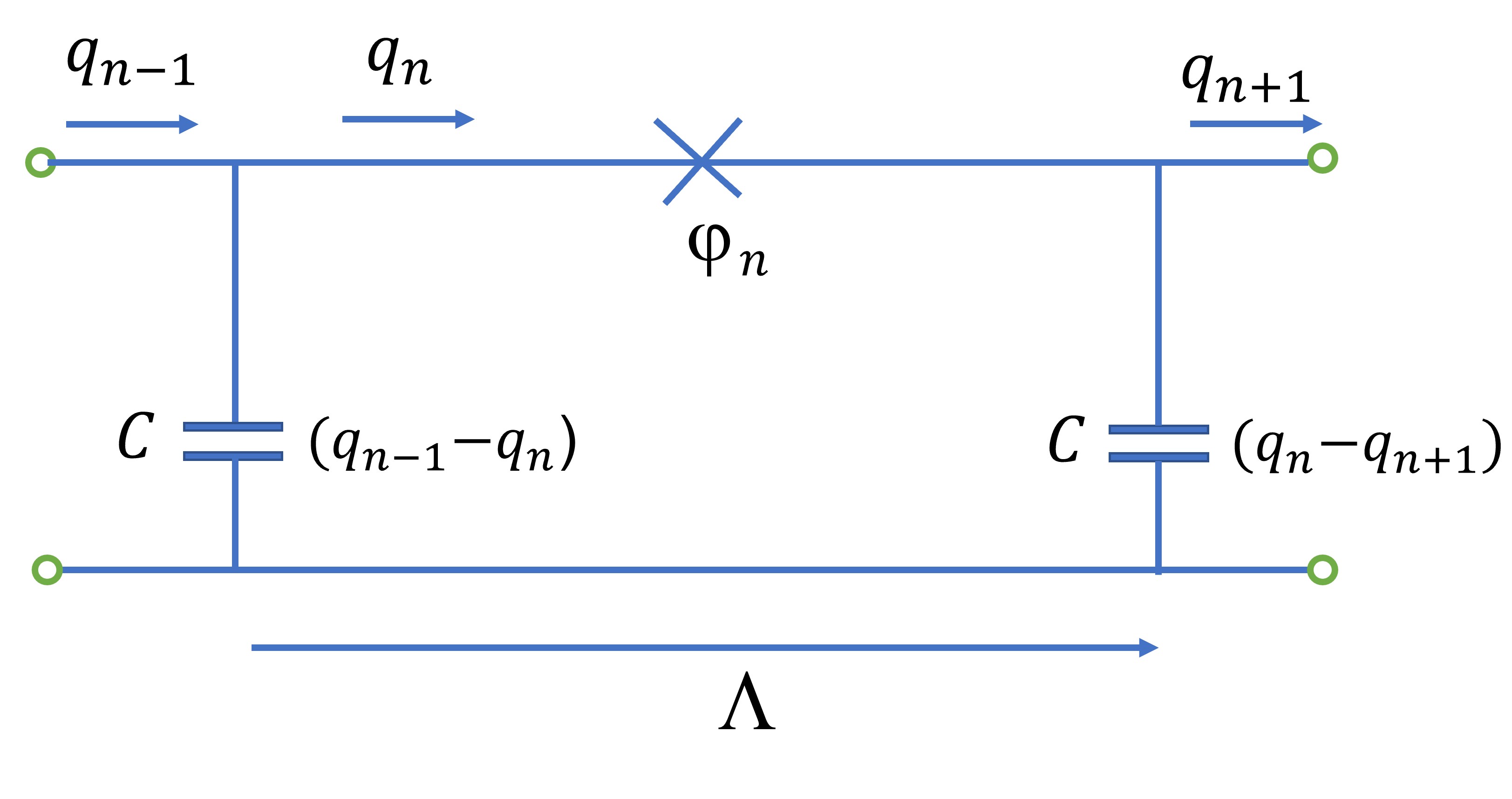}
\vskip -.5cm
\caption{Discrete  JTL. }
 \label{trans1}
\end{figure}

Consider the  model of JTL constructed from identical JJ and capacitors, which is shown on Fig. \ref{trans1}.
We take
as dynamical variables  the phase differences (which we for brevity will call just phases) $\varphi_n$ across the  JJ
and the charges $q_n$ which have passed through the  JJ.
The  circuit equations are
\begin{subequations}
\label{ave7}
\begin{alignat}{4}
\frac{\hbar}{2e}\frac{d \varphi_n}{d t}&=\frac{1}{C}\left(q_{n+1}-2q_{n}+q_{n-1}\right)  \, ,\label{ave7a}\\
\frac{dq_n}{dt} &=   I_c\sin\varphi_n \, ,\label{ave7b}
\end{alignat}
\end{subequations}
where    $C$ is the capacitance, and  $I_c$ is the critical current of the JJ.
Differentiating Eq. (\ref{ave7a}) with respect to $t$ and substituting $dq_n/dt$  from Eq. (\ref{ave7b}), we obtain closed equation for $\varphi_n$
\begin{eqnarray}
\label{com}
\frac{d^2 \varphi_n}{d \tau^2}=\sin\varphi_{n+1}-2\sin\varphi_{n}+\sin\varphi_{n-1}\,,
\end{eqnarray}
where we have introduced the dimensionless time $\tau=t/\sqrt{L_JC}$,
and $L_J=\hbar/(2eI_c)$.

It is interesting to compare Eq. (\ref{com}) with a discretized $\phi^4$ theory \cite{kevrekidis2}
\begin{eqnarray}
\label{po}
\frac{d^2 \phi_n}{d \tau^2}=\frac{1}{h^2}(\phi_{n+1}-2\phi_{n}+\phi_{n-1})+2\phi_n\left(1-\phi_n^2\right)\,,
\end{eqnarray}
and a discrete sine-Gordon equation for lattice wave field \cite{malomed2}
\begin{eqnarray}
\label{po2}
\frac{d^2 \varphi_n}{d \tau^2}-D(\varphi_{n+1}-2\varphi_{n}+\varphi_{n-1})+\sin\varphi_n=0\,.
\end{eqnarray}
where $h$ and $D$ are some constants. Comparing Eqs. (\ref{po}) and (\ref{po2}) with (\ref{com}), one realizes, that for the JTL
 the  non-linearity
enters into the problem in a totally different way. We'll see later that our problem has an additional free parameter -- the amplitude of the wave, which, in particular, opens the way for the controlled perturbation theory.

Let us return to Eq. (\ref{ave7}).
The kinks, we'll be interested in, are localized and characterised by the boundary conditions
\begin{eqnarray}
\label{granu}
\lim_{n\to -\infty}\varphi=\varphi_2\;,\hskip 1cm
\lim_{n\to +\infty}\varphi=\varphi_1\,.
\end{eqnarray}
Summing up  (\ref{ave7a}) from far to the left of the kink up to far to the right of the kink we obtain
\begin{eqnarray}
\label{ansatz}
\frac{\hbar}{2e}\frac{d}{dt}\sum_n\varphi_n
=\frac{1}{C}\left[\left(q_{n+1}-q_{n}\right)_1-\left(q_{n+1}-q_{n}\right)_2\right]  \,.
 \end{eqnarray}

Further on in this paper,  instead of the index $n$ will use a continuous variables $z=n\Lambda$
and will be mostly interested in the running wave solutions  of the form
\begin{eqnarray}
\label{run}
\varphi(z,t)=\varphi(x)\,,\hskip 1cm q(z,t)=q(x)\,,
\end{eqnarray}
where $x=Ut-z$, and $U$ is the running wave velocity.
The boundary conditions become
\begin{eqnarray}
\label{gran}
\lim_{x\to -\infty}\varphi=\varphi_1\;,\hskip 1cm
\lim_{x\to +\infty}\varphi=\varphi_2\,.
\end{eqnarray}

From the  running wave ansatz follows
\begin{eqnarray}
\label{su}
\frac{d}{dt}\sum_n\varphi_n=\frac{U}{\Lambda}\left(\varphi_1-\varphi_2\right)\, .
\end{eqnarray}
To deal with the r.h.s. of (\ref{ansatz})  we need to approximate the finite difference  only far away from the kink, where everything changes slowly, and  the continuum approximation
\begin{eqnarray}
\label{q}
q_{n+1}-q_{n}=\Lambda\frac{\partial q}{\partial z}
\end{eqnarray}
is enough.
From (\ref{q})
and the running wave ansatz follows
\begin{eqnarray}
\label{qu}
\left(q_{n+1}-q_{n}\right)_i
=\frac{\Lambda}{U}\left(\frac{d q_n}{d t}\right)_i=
\frac{\Lambda}{U}\sin \varphi_i\,.
\end{eqnarray}
Substituting (\ref{su}) and (\ref{qu}) into (\ref{ansatz})  we get
for the running wave velocity
\begin{eqnarray}
\label{velocity}
\overline{U}^2=\frac{\sin\varphi_1-\sin\varphi_2}{\varphi_1-\varphi_2}
\equiv \overline{U}^2_{\text{sh}}(\varphi_1,\varphi_2)\,.
\end{eqnarray}
In this paper, for any velocity $V$, $\overline{V}\equiv V\sqrt{L_JC}/\Lambda$.
The reason, why we have chosen subscript sh for the velocity in (\ref{velocity}), will become clear in Section~\ref{shunti}.

To find the profile of the running wave we have to approximate the finite difference in the r.h.s. of (\ref{ave7a}) everywhere, including the regions where the variables change fast.
We can write down (at least formally) the infinite Taylor expansion
\begin{eqnarray}
\label{comm33}
q_{n+1}-2q_{n}+q_{n-1}
=\Lambda^2\frac{\partial^2q}{\partial z^2}+\frac{\Lambda^4}{12}\frac{\partial^4q}{\partial z^4}+\dots\,.
\end{eqnarray}

For the running waves, substituting into the r.h.s. of (\ref{comm33}) the derivative of $q$ with respect to $z$ from (\ref{ave7b}) and then substituting the result into  (\ref{ave7a}), we obtain the ordinary differential equation
\begin{eqnarray}
\label{v7}
\overline{U}^2\frac{d\varphi}{d x}=\frac{d\sin\varphi}{d x}+\frac{\Lambda^2}{12}\frac{d^3\sin\varphi}{d x^3}+\dots\,.
\end{eqnarray}
Integrating with respect to $x$
we obtain
\begin{eqnarray}
\label{v90}
\frac{\Lambda^2}{12}\frac{d^2\sin\varphi}{d x^2}+\dots
=-\sin\varphi +\overline{U}^2\varphi+F\,,
\end{eqnarray}
where $F$ is the constant of integration.
Substituting (\ref{gran}) into (\ref{v90}) we obtain
\begin{eqnarray}
\label{v797}
-\sin\varphi_i +\overline{U}^2\varphi_i+F=0\,,\hskip 1cm i=1,2\,.
\end{eqnarray}
Solving (\ref{v797}) relative to $\overline{U}^2$ and $F$
we recover (\ref{velocity}) and also obtain
\begin{eqnarray}
\label{vb}
F = \frac{\varphi_1\sin\varphi_2-\varphi_2\sin\varphi_1}{\varphi_1-\varphi_2}\,.
\end{eqnarray}

\section{The elasticity theory: the kinks and the solitons}
\label{general}

Now we make the  assumption, by keeping in Eq. (\ref{comm33}) only the first three terms
\begin{eqnarray}
\label{comm33b}
q_{n+1}-2q_{n}+q_{n-1}
=\Lambda^2\frac{\partial^2q}{\partial z^2}+\frac{\Lambda^4}{12}\frac{\partial^4q}{\partial z^4}+\frac{\Lambda^6}{360}\frac{\partial^6q}{\partial z^6}.
\end{eqnarray}
We will call (\ref{comm33b}) the quasi-continuum approximation.
We have seen above that the terms with the derivatives higher than the second are necessary to obtain a physically meaningful results. On the other hand, the term with the
 6th derivative is necessary so the Eq. (\ref{com}) after the truncation
\begin{eqnarray}
\label{old}
\frac{\partial^2 \varphi}{\partial \tau^2}
=\Lambda^2\frac{\partial^2\sin\varphi}{\partial z^2}
+\frac{\Lambda^4}{12}\frac{\partial^4\sin\varphi}{\partial z^4}+\frac{\Lambda^6}{360}\frac{\partial^6 \sin\varphi}{\partial z^6}
\end{eqnarray}
would be non-singular at small wavelengths.

Introducing the notation $y=\sin\varphi$, we write down Eq. (\ref{v90}) after the truncation as
\begin{eqnarray}
\label{v99}
\frac{\Lambda^4}{360}y^{(IV)}+\frac{\Lambda^2}{12}y''=-y +\overline{U}^2\sin^{-1}y+F\,.
\end{eqnarray}
We recognize the equation of equilibrium of  bent and compressed rods for the case of small deflections
\cite{landau}, $\Lambda^4/360$ playing the role of the bending modulus and $\Lambda^2/12$ playing the role of the compressing force. The rod is placed in the external force field,
described alternatively by the potential energy $\Pi(y)$ given by
(\ref{v10b}).

One important feature of the solutions  of (\ref{v99}) can be seen without solving the equation: the localized  solutions at the infinite line with the boundary conditions (\ref{gran}) and the finite energy exist only if
\begin{eqnarray}
\label{ph}
\varphi_2=\pm\varphi_1\,.
\end{eqnarray}
If $\varphi_2=-\varphi_1$ we can talk about the kinks, if $\varphi_2=\varphi_1$ -- about the solitons.

In fact, we know that
the solutions of (\ref{v99}) may be obtained from the variational principle \cite{landau}. We have to make stationary the functional
\begin{eqnarray}
\label{rod}
F_{\text{rod}}=\frac{\Lambda^4}{720}\int y''^2dx-\frac{\Lambda^2}{24}\int y'^2dx
+\int\Pi(y)dx\,.
\end{eqnarray}
The variational principle being formulated, we immediately understand the necessity of the relation
\begin{eqnarray}
\label{phii}
\Pi(\varphi_1)=\Pi(\varphi_2),
\end{eqnarray}
Otherwise, by shifting the kink or the soliton we can change the functional linearly with respect to the shift. Combining (\ref{phii}) with
(\ref{v797}) we obtain (\ref{ph}).

\section{Newtonian equation: the kinks and the   solitons}
\label{shock}

Let us simplify Eq. (\ref{comm33b}) to
\begin{eqnarray}
\label{comm3}
q_{n+1}-2q_{n}+q_{n-1}
=\Lambda^2\frac{\partial^2q}{\partial z^2}+\frac{\Lambda^4}{12}\frac{\partial^4q}{\partial z^4}.
\end{eqnarray}
We will call (\ref{comm3}) the reduced quasi-continuum approximation and
will see later that in certain limiting cases it  can be rigorously justified.
After the simplification, Eq. (\ref{v99}) reduces to
\begin{eqnarray}
\label{v9}
\frac{\Lambda^2}{12}\frac{d^2\sin\varphi}{d x^2}
=-\sin\varphi +\overline{U}^2\varphi+F\,.
\end{eqnarray}
We can consider $x$ as  time and  $\sin\varphi$ as the coordinate of the fictitious particle, visualizing (\ref{v9}) as  Newtonian equation.
Thus the problem of finding the profile of the kink is reduced to studying
the motion of the particle which
starts from  an equilibrium position,
and ends in an equilibrium position.

Multiplying Eq. (\ref{v9}) by the integrating multiplier $d\sin\varphi/dx$ and   integrating once again
 we obtain
\begin{eqnarray}
\label{v10}
\frac{\Lambda^2}{24}\left(\frac{d\sin \varphi}{d x}\right)^2+\Pi(\sin\varphi)=E\,,
\end{eqnarray}
where
\begin{eqnarray}
\label{v10b}
\Pi(\sin\varphi)=\frac{1}{2}\sin^2\varphi-\overline{U}^2(\varphi\sin\varphi
+\cos\varphi)-F\sin\varphi\,,\nonumber\\
\end{eqnarray}
and  $E$ is another constant of integration.
Using the expertise we acquired in mechanics classes,  we come to the conclusion that   the initial  position
corresponds to  maxima  of the "potential energy" $\Pi(\sin\varphi)$, and so does the final position. Note that from the energy conservation law we recover (\ref{phii}) and, hence, (\ref{ph}).

One should compare the kink velocity  with the velocity $u(\varphi)$ of propagation along the  JTL of  small amplitude smooth disturbances of phase on a homogeneous background $\varphi$  \cite{kogan}
\begin{eqnarray}
\label{veveve}
\overline{u}^2(\varphi)=\cos\varphi
\end{eqnarray}
(in this paper we consider only the solutions which lie completely in the sector  $(-\pi/2,\pi/2)$.)
From the fact that there is a maximum of the "potential energy" at the points $\varphi_{1,2}$, follows that
\begin{eqnarray}
\left.\frac{d^2\Pi(\varphi)}
{d\varphi^2}\right|_{\varphi=\varphi_{1,2}}<0\,.
\end{eqnarray}
Calculating the derivatives we obtain
\begin{eqnarray}
\overline{U}^2>\cos\varphi_{1,2}\,,
\end{eqnarray}
that is the running wave is supersonic.

\begin{figure}[h]
\includegraphics[width=.7\columnwidth]{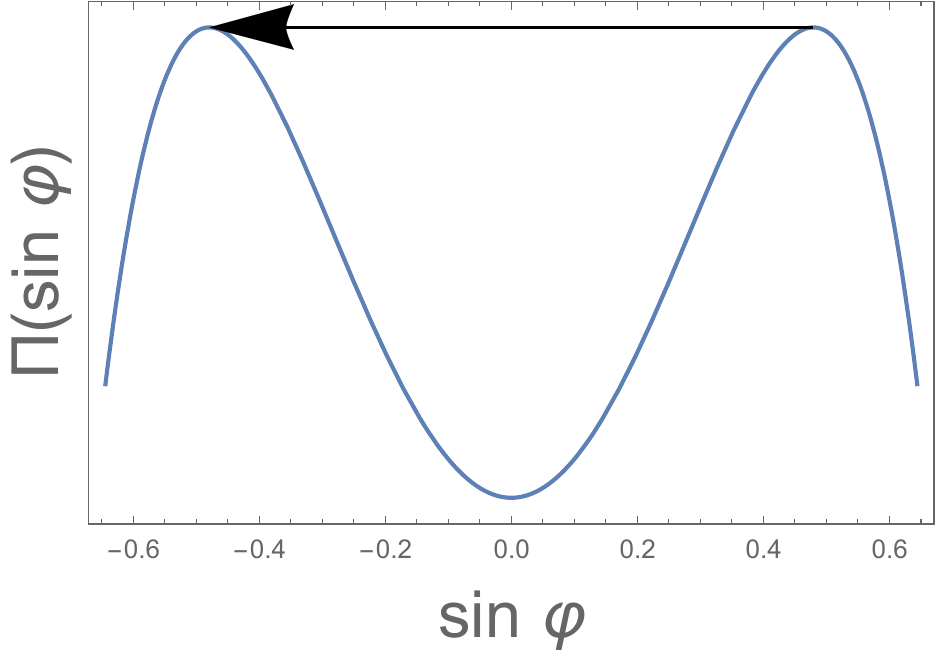}
\vskip .3cm
\includegraphics[width=.7\columnwidth]{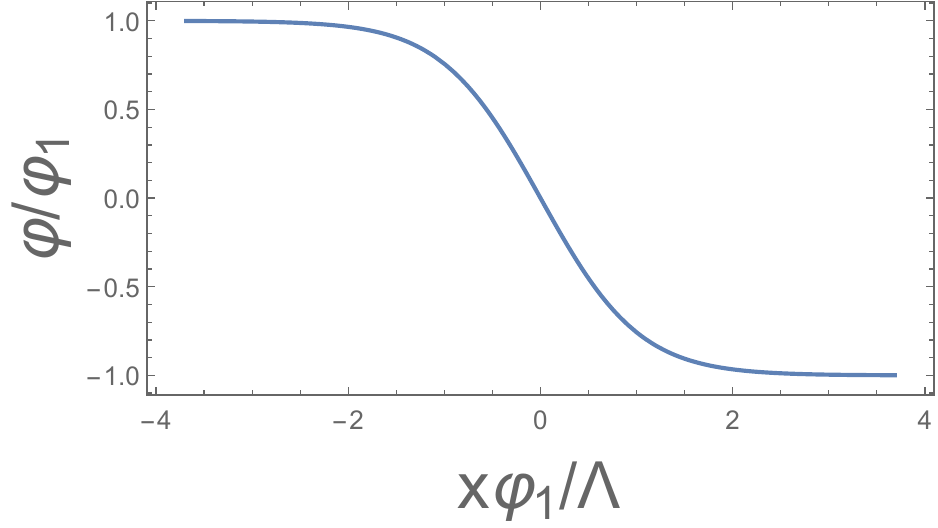}
\caption{The "potential energy" (\ref{v100}) (above) and the  kink profile calculated  with this energy according to Eq. (\ref{v10}) (below). We have chosen $\varphi_1=.5$.}
 \label{trans2}
\end{figure}

Adding the energy conservation law to  (\ref{v797})
we obtain
\begin{subequations}
\label{ss}
\begin{alignat}{4}
F&=0 \, ,\label{felo}\\
\overline{U}^2&=\overline{U}^2_{\text{sh}}(\varphi_1,-\varphi_1)=\frac{\sin\varphi_1}{\varphi_1} \equiv\overline{U}^2_{\text{k}}(\varphi_1)\,,\label{velo}
\end{alignat}
\end{subequations}
and, after  the substitution into (\ref{v10b}),
\begin{eqnarray}
\label{v100}
\Pi(\sin\varphi)=\frac{1}{2}(\sin\varphi-\sin\varphi_1)^2 \nonumber\\
-\frac{\sin\varphi_1}{\varphi_1}
[\cos\varphi-\cos\varphi_1-(\varphi_1-\varphi)\sin\varphi]
\end{eqnarray}
(and $E=0$).
The "potential energy" (\ref{v100})  is graphically  presented on Fig. \ref{trans2} (above), and the kink profile --  on Fig. \ref{trans2} (below).

\begin{figure}[h]
\includegraphics[width=.7\columnwidth]{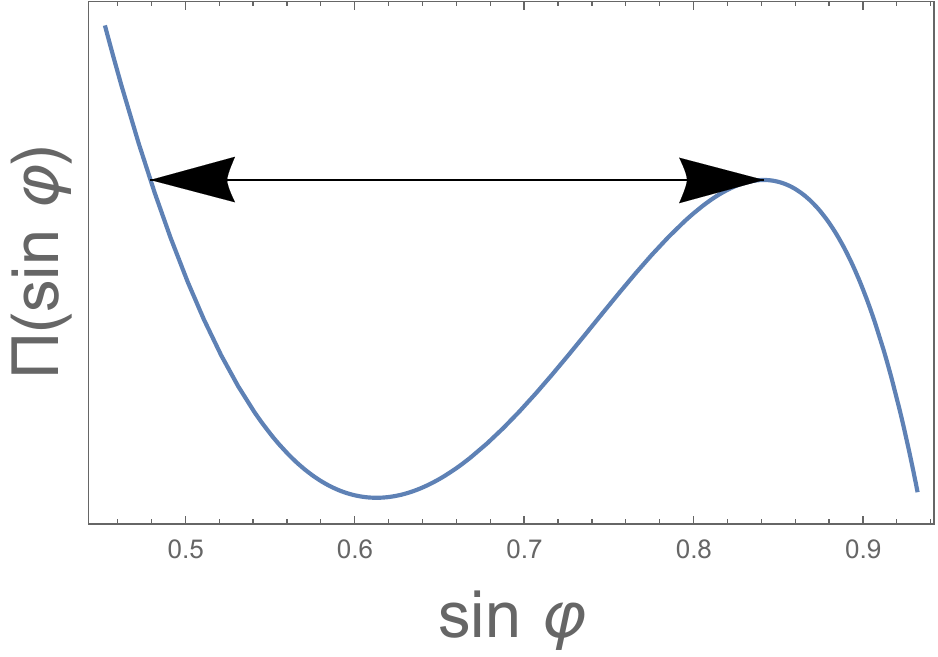}
\vskip .3cm
\includegraphics[width=.7\columnwidth]{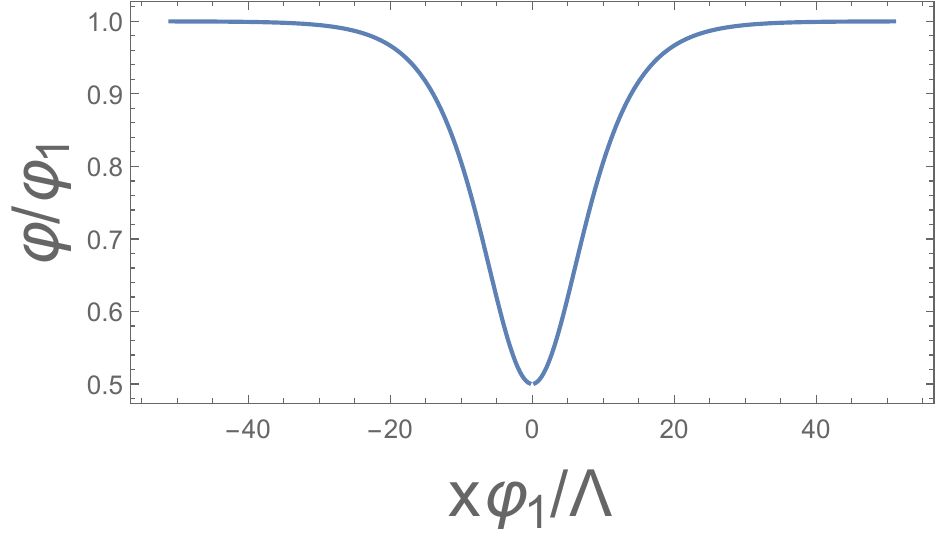}
\caption{The "potential energy" (\ref{v2}) (above) and the  soliton profile
 according to Eq. (\ref{v10}) (below).  We have chosen  $\varphi_1=1.$ and  $\varphi_0=.5$. }
 \label{trans3}
\end{figure}

For the case of the soliton, the two maxima of the potential energy mentioned after Eq. (\ref{v10b}) are the same maximum,
that is the particle
returns to the initial  position after reflection from a potential wall (see Fig.  \ref{trans3}).
Note that due to exactly the same reasons as given in the previous Section for the kink, the soliton is also supersonic.
In this case the two equations of (\ref{v797}) become one equation.
As an
additional parameter we take the amplitude of the soliton  (maximally different from $\varphi_1$ value of $\varphi$), which we will  designate as $\varphi_0$. Adding to (\ref{v797}) the equation
\begin{eqnarray}
\label{pio}
\Pi(\sin\varphi_0) = \Pi(\sin\varphi_1)
\end{eqnarray}
and solving the obtained system  we obtain
\begin{subequations}
\label{ss2}
\begin{alignat}{4}
\overline{U}^2_{\text{sol}}(\varphi_1,\varphi_0)
&=\frac{\left(\sin\varphi_1-\sin\varphi_0\right)^2}
{2[\cos\varphi_0-\cos\varphi_1-(\varphi_1-\varphi_0)\sin\varphi_0]}\,,\label{velo2}\\
\Pi(\sin\varphi)&=\frac{1}{2}\left(\sin\varphi_1-\sin\varphi\right)^2
-\overline{U}^2_{\text{sol}}(\varphi_1,\varphi_0)\nonumber\\
&\cdot\left[\cos\varphi-\cos\varphi_1-(\varphi_1-\varphi)\sin\varphi\right] \label{v2}
\end{alignat}
\end{subequations}
(and $E=0$). Note that while derivation of the formula for the kink velocity
demands approximation of the wave behavior only far away from the kink, derivation of the formula for the soliton velocity  demands approximation of the wave behaviour in the region of the soliton.
The "potential energy" (\ref{v2}) is graphically  presented on Fig. \ref{trans3} (above), and the soliton profile -- on Fig. \ref{trans3} (below).

\section{Weak kinks and weak solitons}
\label{weak}

Consider specifically  the limiting case of weak kinks ($|\varphi_1|\ll 1$).
Expanding the "potential energy" with respect to $\varphi$ and $\varphi_1$  and keeping only the lowest order terms we obtain the approximation to Eq. (\ref{v10}) in the form
\begin{eqnarray}
\label{v101}
\Lambda^2\left(\frac{d \varphi}{d x}\right)^2=\left(\varphi_1^2-\varphi^2\right)^2\,.
\end{eqnarray}
The solution of Eq. (\ref{v101})  is
\begin{eqnarray}
\label{v11}
\varphi(x)=-\varphi_1\tanh \frac{\varphi_1x}{\Lambda}\,.
\end{eqnarray}
Equations (\ref{v11}) coincides with that obtained by Katayama et al. \cite{katayama}.
So does
Eq. (\ref{velo}), being expanded in series with respect to $\varphi_1$ and truncated after the first two terms:
\begin{eqnarray}
\label{velok}
\overline{U}_k^2(\varphi_1)=1-\frac{\varphi_1^2}{6}\,.
\end{eqnarray}

In the limiting  case of weak solitons  ($\Delta\varphi\tan\varphi_1\ll 1$, $\varphi_1\sim 1$, where $\Delta\varphi\equiv \varphi_1-\varphi_0$), it is convenient to make the change of variable $\psi\equiv\varphi-\varphi_1$, after which Eq. (\ref{v10}) takes the form
\begin{eqnarray}
\label{sol3}
\Lambda^2\left(\frac{d \psi}{d x}\right)^2
=4\tan\varphi_1\cdot\psi^2
\left(\psi+\Delta\varphi\right)\,.
\end{eqnarray}
The solution of Eq. (\ref{sol3}) is
\begin{eqnarray}
\label{v14}
\psi=-\Delta\varphi\sech^2\left( \sqrt{\Delta\varphi\tan\varphi_1} x/\Lambda\right)\,.
\end{eqnarray}
Velocity of the soliton in this approximation is
\begin{eqnarray}
\label{velok2}
\overline{U}_{\text{sol}}^2(\varphi_1,\varphi_0)
=\cos\varphi_1\left(1+\frac{\tan\varphi_1}{3}\Delta \varphi\right).
\end{eqnarray}

Looking at Eqs. (\ref{v11})  and (\ref{v14}) we realize with the hindsight that the reduced quasi-continuum approximation
can be rigorously justified when the  running wave is a small perturbation on a homogeneous background.  Actually, the equations say more than that.
Common wisdom says that the continuum approximation and the small amplitude approximation  are independent - there could be a wave with small amplitude, which allows to expand the sine function, but which varies fast in space (wavelength comparable to lattice spacing), so the continuum limit is not justified. And there could be the opposite situation (large amplitude, long wavelength), in which the sine needs to be retained but the continuum limit is allowed.

However, for the running waves in the discrete JTL these approximations are not independent. Parametrically, the length scale of the waves is of the
order of the  lattice spacing $\Lambda$, so, naively, the  quasi-continuum approximation can not be justified. What we have shown above, is that for the
weak kinks the length scale is $\Lambda/|\varphi_1|$, and for the weak solitons the length scale is
$\left(\Lambda/\sqrt{|\Delta\varphi|}\right)$, thus justifying the reduced quasi-continuum approximation in both cases.

\section{The shocks}
\label{shunti}

\begin{figure}[h]
\includegraphics[width=\columnwidth]{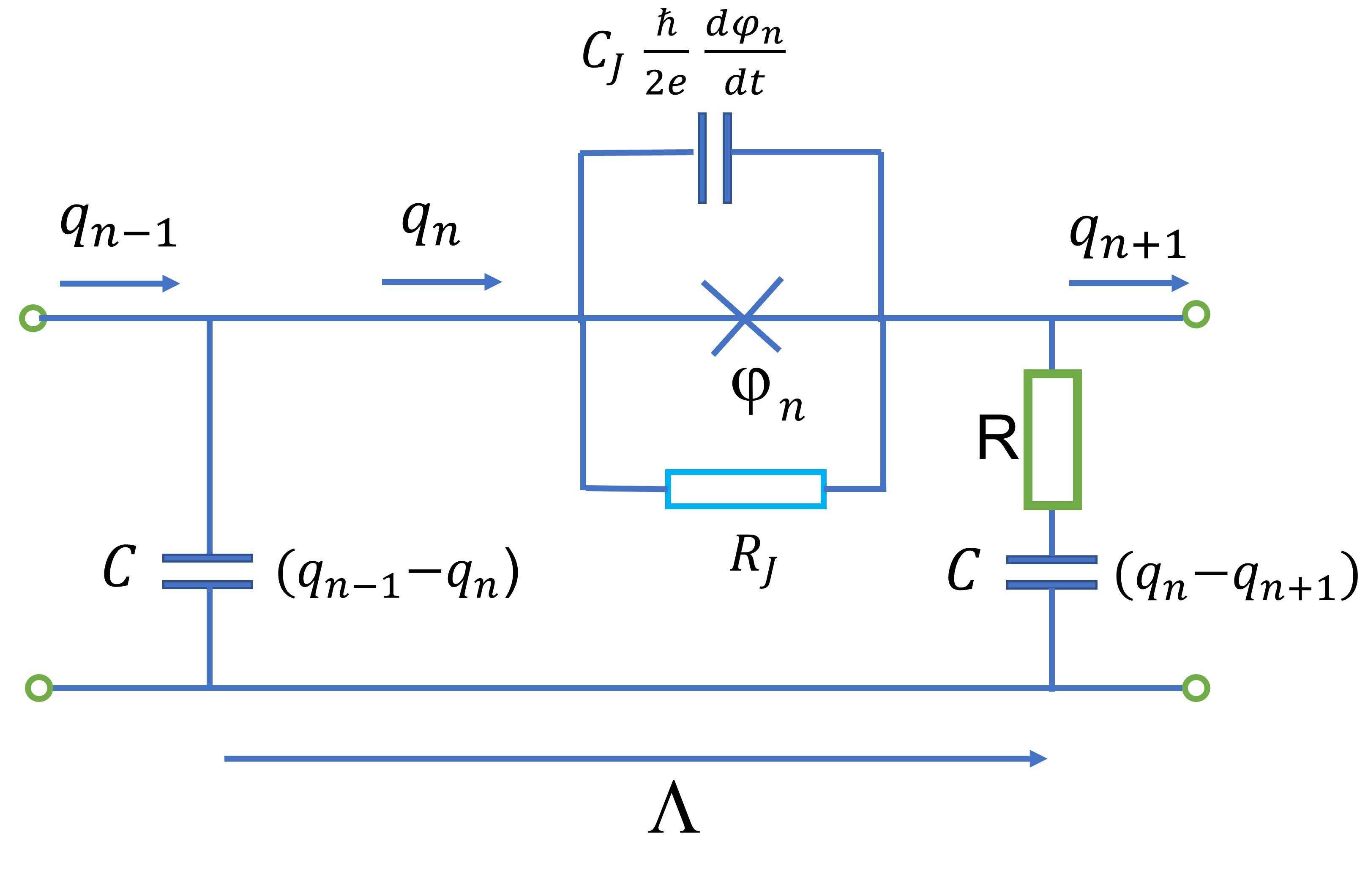}
\caption{Discrete JTL with the capacitor  and the resistor shunting the JJ and another resistor in series with the ground capacitor}
 \label{trans4}
\end{figure}

Consider JTL with the capacitor  and resistor shunting the JJ and another resistor in series with the ground capacitor,
shown  on Fig. \ref{trans4}.
 As the result,
 Eq. (\ref{ave7}) changes to
\begin{subequations}
\label{ave8}
\begin{alignat}{4}
\frac{\hbar}{2e}\frac{d \varphi_n}{d t}&=\left(\frac{1}{C}+R\frac{\partial}{\partial t}\right)\left(q_{n+1}-2q_{n}+q_{n-1}\right)  \, ,\label{ave8a}\\
\frac{dq_n}{dt} &=   I_c\sin\varphi_n+\frac{\hbar}{2eR_J}\frac{d \varphi_n}{d t}
+C_J\frac{\hbar}{2e}\frac{d^2\varphi_n}{d t^2}\, ,\label{ave8b}
\end{alignat}
\end{subequations}
where $R$ is the ohmic resistor  in series with the ground
capacitor, and $C_J$ and $R_J$ are the capacitor and the ohmic resistor shunting the JJ.

Considering again the running wave solutions we obtain the generalization of Eq. (\ref{v9})
\begin{eqnarray}
\label{v9b}
\frac{\Lambda^2}{12}\frac{d^2\sin\varphi}{d x^2}
+\left(\frac{C_J}{C}+\frac{R}{R_J}\right)\overline{U}^2\Lambda^2
\frac{d^2\varphi}{d x^2}\nonumber\\
+\left(\frac{R}{Z_J}\cos\varphi+\frac{Z_J}{R_J}\right)\overline{U}
\Lambda\frac{d\varphi}{d x}
=-\sin\varphi+\overline{U}^2\varphi+F\,,
\end{eqnarray}
where $Z_J\equiv\sqrt{L_J/C}$ is the characteristic impedance of the JTL, and  we discarded the terms with the derivatives higher than of the forth order.

We impose  the boundary conditions (\ref{gran}) and try to understand
what part of the analysis of Section~\ref{shock} can be transferred to the present case.
The results  (\ref{v797})
are determined only by the r.h.s. of Eq. (\ref{v9}),  so are (\ref{v7}), following from (\ref{v797}).
Since the r.h.s. of Eqs. (\ref{v9}) and (\ref{v9b}) are identical, these equations are valid in the present case also. In particular, we obtain
\begin{eqnarray}
\overline{U}^2=\overline{U}^2_{\text{sh}}(\varphi_1,\varphi_2) \, .
\end{eqnarray}
We emphasise that the velocity of the shock wave
does not depend upon the dissipation, similar to the case of KdV equation \cite{jarmo}, but in distinction to the case of nonlinear Schrodinger equation \cite{cai}.

On the other hand, the resistors, by introducing the effective "friction force", break the "energy" conservation law, which means that the stationary kinks and the solitons we considered previously are no longer possible, however weak the dissipation is. However in the lossy JTL the solutions  with $|\varphi_2|\neq |\varphi_1|$ (the shocks) are possible.

\subsection{The qualitative analysis}
\label{qa}

We saw in Section~\ref{shock} that if
\begin{eqnarray}
\label{p}
\left(\frac{C_J}{C}+\frac{R}{R_J}\right)\overline{U}^2\ll 1\,,
\end{eqnarray}
 Eq. (\ref{v9b}) can be reduced to Newtonian form.
The situation is even simpler when the inequality (\ref{p}) is inverted. In this case the
first term in the l.h.s. of (\ref{v9b}) can be neglected, and the equation is already in Newtonian form. In the latter case the discrete nature of the JTL doesn't manifest itself --the continuum approximation is valid \cite{kogan}. In each of these cases,
the fictitious particle motion describing the shock connects the "potential energy" maximum at $\varphi=\varphi_1$ with the "potential energy" minimum at $\varphi=\varphi_2$.

For qualitative analysis  of  (\ref{v9b}) when the first two terms in the l.h.s. of the equation are comparable, it is better to
present it  as a system of two first order differential equations
\begin{subequations}
\label{system}
\begin{alignat}{4}
\left[\frac{\cos\varphi}{12}
+\left(\frac{C_J}{C}+\frac{R}{R_J}\right)\overline{U}^2\right]
\Lambda\frac{d\chi}{d x}=\frac{\sin\varphi}{12}\chi^2\nonumber\\
-\left(\frac{R}{Z_J}\cos\varphi+\frac{Z_J}{R_J}\right)\overline{U}\chi
-\sin\varphi +\overline{U}^2\varphi+F\,,  \\
\Lambda\frac{d\varphi}{dx}=\chi\,,
\end{alignat}
\end{subequations}

Now, one important feature of  shocks  can be understood immediately. We are talking about the direction of shock propagation.
Linearising Eq. (\ref{system}) in the vicinity of the fixed points
 $(\chi,\varphi)=(0,\varphi_1)$ and $(\chi,\varphi)=(0,\varphi_2)$ we obtain
\begin{eqnarray}
\label{ht}
\Lambda\left(\begin{array}{l}d\chi/d x\\ d\varphi/dx\end{array}\right)
=\left(\begin{array}{cc}M_i & N_i\\1 & 0\end{array}\right)
\left(\begin{array}{c}\varphi-\varphi_i\\\chi\end{array}\right),\;\;i=1,2
\end{eqnarray}
where
\begin{subequations}
\begin{alignat}{4}
M_i&=-\left(\frac{R}{Z_J}\cos\varphi_i+\frac{Z_J}{R_J}\right)\overline{U}\,,\\
N_i&=\frac{\overline{U}^2-\cos\varphi_i}{\cos\varphi_i/12
+\left(C_J/C+R/R_J\right)\overline{U}^2}\,.
\end{alignat}
\end{subequations}
The eigenvalues of the matrix in (\ref{ht}) are
\begin{eqnarray}
\label{lam}
\lambda_{i,\pm}=\frac{M_i\pm\sqrt{M_i^2+4N_i}}{2}\,.
\end{eqnarray}
Thus negative $N_i$ corresponds to a stable fixed point, and positive $N_i$ -- to a semi-stable fixed point.
From the fact  that $\varphi_1$ is a semi-stable fixed point, and $\varphi_2$ is a stable fixed point we obtain
\begin{eqnarray}
\label{un22}
\cos\varphi_2>\overline{U}^2_{\text{sh}}(\varphi_1,\varphi_2)
>\cos\varphi_1\,.
\end{eqnarray}

The inequalities (\ref{un22}) allow only one direction of shock propagation
- from smaller $\cos\varphi$ to larger $\cos\varphi$.
Taking into account (\ref{veveve}), we can present (\ref{un22}) as
\begin{eqnarray}
\overline{u}^2(\varphi_2)>\overline{U}^2_{\text{sh}}(\varphi_1,\varphi_2)
>\overline{u}^2(\varphi_1)\,,
\end{eqnarray}
thus establishing the connection with the  well known  in the nonlinear waves theory  fact: the shock velocity is higher than the sound velocity   in the region before the shock
but lower than the sound velocity in the region behind the shock  \cite{whitham}.

\begin{figure}[h]
\includegraphics[width=.65\columnwidth]{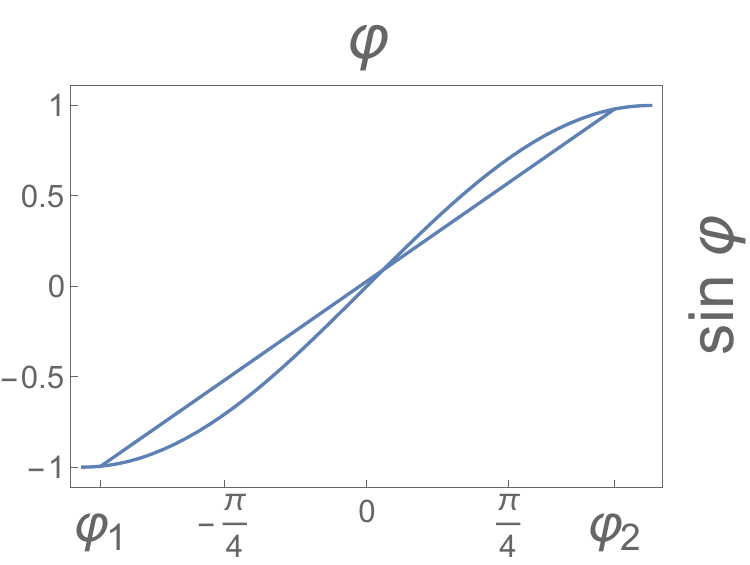}
\vskip .5cm
\includegraphics[width=.65\columnwidth]{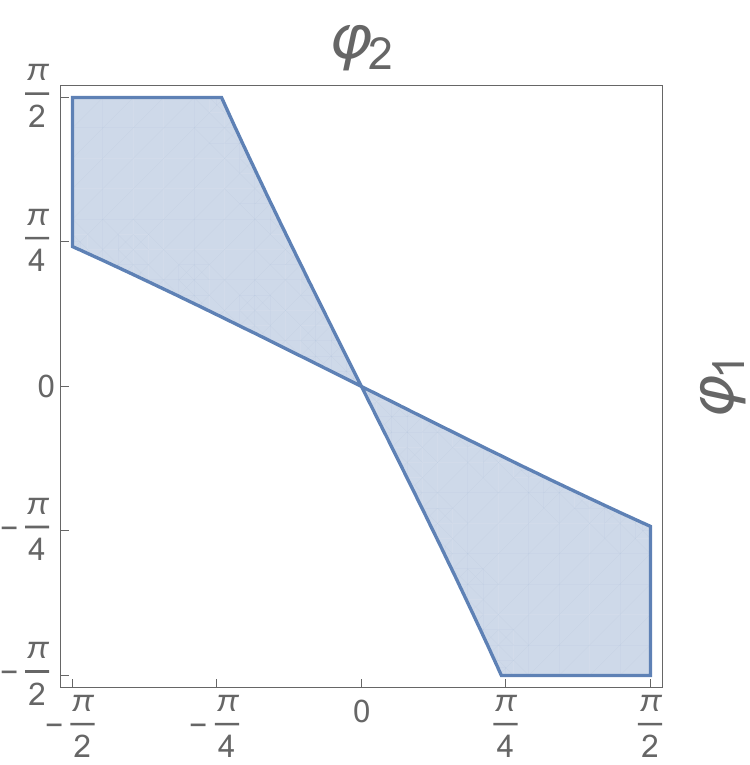}
\caption{(above): The geometric property of the points belonging to the shaded region.
(below): The phase space of the boundary conditions on the ends of the JTL $\varphi_1$ and $\varphi_2$. The region, which corresponds to the forbidden shock boundary conditions, is shaded.}
\label{trans10}
\end{figure}

Let us write down inequalities (\ref{un22}) explicitly
\begin{eqnarray}
\label{inequality3}
\cos\varphi_2>\frac{\sin\varphi_1-\sin\varphi_2}{\varphi_1-\varphi_2}
>\cos\varphi_1\,.
\end{eqnarray}
We will combine the case we studied up to now, when   $\varphi_1$ was the phase before the shock and $\varphi_2$ - behind the shock, with the opposite case, which corresponds to
 indices 1 and 2 in (\ref{inequality3}) being interchanged.
The points in the phase space of the shock boundary conditions $(\varphi_1,\varphi_2)$, for which neither (\ref{inequality3}),
nor its interchanged version
 are  satisfied, and hence the shock is forbidden, can be visualized by the fact that the secant of the curve $\sin\varphi$ between the points  crosses the curve, like it is shown on Fig. \ref{trans10} (above).
Because $\sin\varphi$ is concave downward for
$0<\varphi<\pi/2$, and concave upward for $-\pi/2<\varphi<0$, the shock is allowed between any pair of
$\varphi_1,\varphi_2$ having the same sign.  For $\varphi_1$ and $\varphi_2$ having opposite signs the shock may be allowed or not.
We  present the phase space of shock boundary conditions  on Fig. \ref{trans10} (below).
The forbidden region is shaded.

When the asymptotic phases on the two sides of the JTL belong to the shaded region,
probably the forbidden shock is split into two allowed ones:   between $\varphi_{1}$ and  some intermediate $\varphi_{in}$, and between $\varphi_{2}$ and $\varphi_{in}$.   Say, when   $\varphi_2=-\varphi_1$, the system may chose the intermediate value $\varphi_{in}=0$. In this hypothetical case, the shocks move in the opposite directions, and the central part with the phase $\varphi_{in}=0$ expands with the velocity $2U_{\text{k}}(\varphi_1,0)$. However, the case of multiple shocks, being simultaneously present
in the system,  demands further studies.

\subsection{The numerical integration}

\begin{figure}[h]
\includegraphics[width=.8\columnwidth]{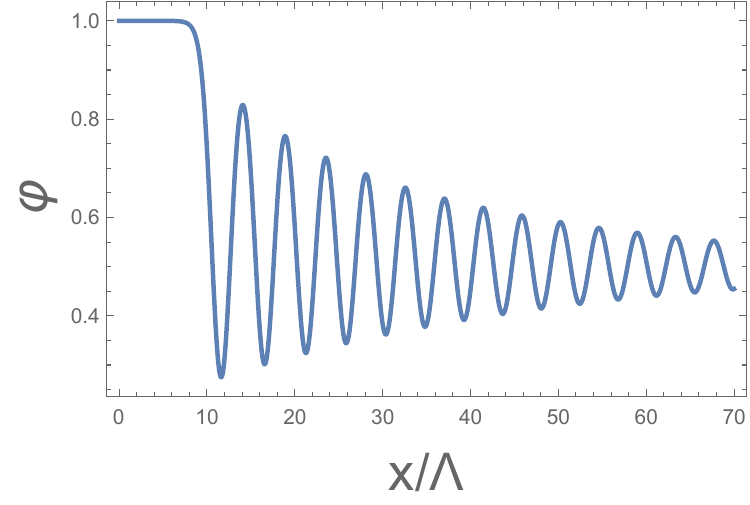}
\caption{The shock profile  according to Eq. (\ref{system2}). We have chosen $\varphi_1=1$, $\varphi_2=.5$, $Z_J/R_J=.005$.}
 \label{trans11}
\end{figure}

Equation (\ref{v9b}) can  be easily integrated numerically in the general case. For aesthetical reasons let us
simplify it by putting $R=0$ and $C_J=0$. (Actually, the physical meaning and the relevance of the resistor in series with the ground
capacitor is not obvious. We included it because we were able to do it for free. The capacitance of the JJ is certainly physically relevant. Anyhow, when $C_J/C\ll 1$, it
can be ignored.) After the simplification and substitution of the results for $\overline{U}$ and $F$ from (\ref{velocity}) and (\ref{vb}),
the equation becomes
\begin{eqnarray}
\label{system2}
&&\frac{\Lambda^2}{12}\frac{d^2\sin\varphi}{d x^2} +\frac{Z_J}{R_J}\overline{U}_{\text{sh}}(\varphi_1,\varphi_2)\Lambda\frac{d\varphi}{dx}=\\
&&\frac{(\varphi-\varphi_2)(\sin\varphi_1-\sin\varphi)-
(\sin\varphi-\sin\varphi_2)(\varphi_1-\varphi)}{\varphi_1-\varphi_2}.\nonumber
\end{eqnarray}
Note that  for weak shocks ($|\varphi_1-\varphi_2|\ll |\varphi_1|$), the r.h.s. of Eq. (\ref{system2}) simplifies to
\begin{eqnarray}
\label{system3}
\dots
=-\frac{1}{2}\sin\varphi_1(\varphi_1-\varphi)(\varphi-\varphi_2)\,.
\end{eqnarray}
The  result of the numerical integration of (\ref{system2}) is shown on Fig. \ref{trans11} (compare  with Figs. \ref{trans2} (below)).

Dissipation is always present in real experiments. And yet we can observe solitary waves (though they are nonstationary, but practically identical to the corresponding stationary solitons at any given moment of time) in case if dissipation is weak enough. Thus, weak dissipation does not completely kill solitary waves, it just makes them nonstationary/attenuating. Such solitary waves are observed in numerical calculations and in experiments, as was the case with granular chains \cite{nesterenko0,nesterenko}. Also looking at Fig. 6 we realize that weak dissipation results in the oscillatory shock profile demonstrating significance of dispersion in this specific case.  On the other hand, there is a critical rate of dissipation which transforms oscillating stationary shock waves into monotonous as was the case with granular chains \cite{nesterenko3}. This also can be seen from Eq. (\ref{lam}).

\section{The simple wave approximation}
\label{contr}

Though the main subject of the present paper is the running waves,   it is worth to have a look at what happens, when we discard  the running wave ansatz. To obtain tangible results,
let us  modify the quasi-continuum approximation  (\ref{comm33b}) in the following way
\begin{eqnarray}
\label{commb}
q_{n+1}-2q_n+q_{n-1}=\left(\Lambda\frac{\partial}{\partial z}+\frac{\Lambda^3}{24}\frac{\partial^3}{\partial z^3}\right)^2q\,.
\end{eqnarray}
After we apply the modified quasi-continuum approximation to the r.h.s. of Eq. (\ref{ave7a}) and combine thus obtained equation with (\ref{ave7b}),
Eq. (\ref{old}) is modified to
\begin{eqnarray}
\label{combi}
\frac{\partial^2 \varphi}{\partial \tau^2}=\left(\Lambda\frac{\partial}{\partial z}
+\frac{\Lambda^3}{24}\frac{\partial^3}{\partial z^3}\right)^2\sin\varphi\,.
\end{eqnarray}
This modification opens the way for the simple wave approximation, that is decoupling of the wave equation into two separate equations for the right- and left-going waves.
Such decoupling can be easily done for the linear wave equations, as  it is shown in Appendix~\ref{linear}. Following the pattern,
let us decouple (\ref{combi}) into 2 equations
for $|\varphi|\ll 1$ by brute force   as
\begin{eqnarray}
\label{combi2}
\frac{\partial \varphi}{\partial \tau}=\pm\left(\Lambda\frac{\partial}{\partial z}
+\frac{\Lambda^3}{24}\frac{\partial^3}{\partial z^3}\right)\sqrt{\frac{\sin\varphi}{\varphi}}\varphi\,.
\end{eqnarray}
Taking
\begin{eqnarray}
\label{ksi1}
\sqrt{\frac{\sin\varphi}{\varphi}}=1-\frac{\varphi^2}{12}\,,
\end{eqnarray}
 substituting (\ref{ksi1})  into (\ref{combi}) and
keeping  only the leading terms we obtain  the equations
\begin{eqnarray}
\label{om}
\frac{\partial \varphi}{\partial \tau}
=\pm\left(\Lambda\frac{\partial\varphi}{\partial z}
-\frac{\Lambda}{12}\frac{\partial\varphi^3}{\partial z}
+\frac{\Lambda^3}{24}\frac{\partial^3\varphi}{\partial z^3}\right)\,,
\end{eqnarray}
which are modified Korteweg-de Vries (mKdV) equations.

When  $\varphi=\psi+\varphi_1$ ($|\psi|\ll 1$, $\varphi_1\sim 1$),
we present $\sin\varphi$ (ignoring the constant term)  as
\begin{eqnarray}
\sin\varphi=\cos\varphi_1\left(1-\frac{\tan\varphi_1}{2}\psi\right)\psi\,,
\end{eqnarray}
substitute  into (\ref{combi}), extract the square root, taking into account that as
\begin{eqnarray}
\left(1-\frac{\tan\varphi_1}{2}\psi\right)^{1/2}=1-\frac{\tan\varphi_1}{4}\psi\,,
\end{eqnarray}
and keep  only the leading terms to obtain
\begin{eqnarray}
\label{coco}
\frac{\partial \psi}{\partial \tau}
=\pm\sqrt{\cos\varphi_1}\left(\Lambda\frac{\partial\psi}{\partial z}
-\frac{\tan\varphi_1}{4}\Lambda\frac{\partial\psi^2}{\partial z}
+\frac{\Lambda^3}{24}\frac{\partial^3\psi}{\partial z^3}\right),\nonumber\\
\end{eqnarray}
which is Korteweg-de Vries (KdV) equation.

Equations (\ref{om}) and (\ref{coco}) were derived to solve nonstationary problems,
but they also can be conveniently used for describing the running waves, in which case the equations  take the form (after being integrated once)
\begin{eqnarray}
\label{om2}
\frac{\Lambda^2}{24}\frac{d^2\varphi}{d x^2}
=\left(\overline{U}-1\right)\varphi
+\frac{1}{12}\varphi^3+F\,,
\end{eqnarray}
and
\begin{eqnarray}
\label{coco2}
\frac{\Lambda^2}{24}\frac{d^2\psi}{d x^2}
=\left(\overline{U}/\sqrt{\cos\varphi_1}-1\right)\psi
+\frac{\tan\varphi_1}{4}\psi^2+F\,.
\end{eqnarray}
From the boundary conditions (\ref{gran}) for the kink, and from the boundary condition (\ref{gran}) and the energy conservation law (\ref{pio}) for the soliton, we obtain $F=0$ and
\begin{eqnarray}
\overline{U}_{\text{k}}=1-\frac{\varphi_1^2}{12}\,,
\end{eqnarray}
and
\begin{eqnarray}
\overline{U}_{\text{sol}}=\sqrt{\cos\varphi_1}
\left(1+\frac{\tan\varphi_1}{6}\Delta\varphi\right)\,,
\end{eqnarray}
which coincides (for the approximation used) with (\ref{velok}) and (\ref{velok2}).
Integrating (\ref{om2}) and (\ref{coco2})  we recover
(\ref{v101}) and (\ref{sol3}).

If we return to Eq. (\ref{ave8}), discard the running wave ansatz,
assume  that $R=0$ and $C_J=0$ and
assume the dissipation to be small we obtain instead of  (\ref{combi})
\begin{eqnarray}
\label{comik}
\left(\frac{\partial}{\partial \tau}
-\frac{\sqrt{L_J/C}}{2R_J}\frac{\partial^2}{\partial z^2 }\right)^2\varphi
=\left(\frac{\partial}{\partial z}
+\frac{1}{24}\frac{\partial^3}{\partial z^3}\right)^2\sin\varphi.\nonumber\\
\end{eqnarray}

Extracting root square from both parts of (\ref{comik}) we obtain the generalization of
(\ref{om})
\begin{eqnarray}
\frac{\partial \varphi}{\partial \tau}
=&\pm&\left(\frac{\partial\varphi}{\partial z}
-\frac{1}{12}\frac{\partial\varphi^3}{\partial z}
+\frac{1}{24}\frac{\partial^3\varphi}{\partial z^3}\right)\nonumber\\
&+&\frac{\sqrt{L_J/C}}{2R_J}\frac{\partial^2\varphi}{\partial z^2 }\,,
\end{eqnarray}
and the generalization of
 (\ref{coco})
\begin{eqnarray}
\frac{\partial \psi}{\partial \tau}
=&\pm&\sqrt{\cos\varphi_1}\left(\frac{\partial\psi}{\partial z}
-\frac{\tan\varphi_1}{4}\frac{\partial\psi^2}{\partial z}
+\frac{1}{24}\frac{\partial^3\psi}{\partial z^3}\right)\nonumber\\
&+&\frac{\sqrt{L_J/C}}{2R_J}\frac{\partial^2\psi}{\partial z^2 }\,,
\end{eqnarray}
which is KdV equation with dissipation \cite{whitham}.

\section{Nonlinear dispersion law}
\label{stocks}

Let us return to Eqs. (\ref{ave7}) and (\ref{com}). In this Section and in the next one, in distinction from all the previous Sections,  we will not use any kind of continuum or quasi-continuum approximation. The transmission line will be considered as a discrete one. Our perturbation theory will be constructed following Stocks.
In the linear approximation, when we assume $\sin\varphi_n=\varphi_n$, there
exist periodic solutions of the equation in the form
\begin{eqnarray}
\label{re}
\varphi_n(\tau)=a\cos\theta\,,
\end{eqnarray}
where $\theta=\omega t-k n$.
Taking into account that the  sine functions can be expanded into infinite series,
we can construct  perturbative expansion of the solution of Eq. (\ref{com}) starting from (\ref{re})
\begin{eqnarray}
\label{re2}
\varphi_n(\tau)=a\cos\theta+b\cos 3\theta+d\cos 5\theta+\dots\,.
\end{eqnarray}
We will see that  the amplitude $a$ (more exactly $a^2$) will serve as the expansion parameter.

Substituting (\ref{re2}) into (\ref{com}) and equating the coefficients before the cosines we obtain
\begin{subequations}
\label{zz}
\begin{alignat}{4}
L_JC\omega^2a&=\left(4a-\frac{a^3}{2}+\frac{a^5}{48}-\frac{a^2b}{2}\dots\right)
\sin^2\left(\frac{k}{2}\right)  \, ,\label{zza}\\
9L_JC\omega^2b&=\left(-\frac{a^3}{6}+4b+\frac{a^5}{96}-a^2b+\dots\right)\sin^2\left(\frac{3k}{2}\right) \, ,\label{zzb}\\
\dots \nonumber
\end{alignat}
\end{subequations}
From (\ref{zz}) we obtain (up to the relative order of $a^2$)
\begin{subequations}
\begin{alignat}{4}
&\omega(k;a^2)=\frac{2}{\sqrt{L_JC}}\left(1-\frac{a^2}{16}\right)\sin\left(\frac{k}{2}\right)
\equiv \omega_0(k)+\omega_2(k)a^2\, ,\label{disper22}\\
&\varphi_n(\tau)=a\cos(\omega\tau-k n)
-\frac{a^3}{96}\frac{\left(1+2\cos k\right)^2}{2-\cos k-\cos^2 k}\nonumber\\
&\cdot\cos\left[3(\omega\tau-k n)\right]\,.
\end{alignat}
\end{subequations}

Alternatively, the nonlinear dispersion law can be obtained from  Eq. (\ref{ave7}).
We take into account additionally the capacitors shunting the JJ and thus generalize  (\ref{ave7}) to (compare with (\ref{ave8}))
\begin{subequations}
\label{e8}
\begin{alignat}{4}
\frac{\hbar}{2e}\frac{d \varphi_n}{d t}&=\frac{1}{C}\left(q_{n+1}-2q_{n}+q_{n-1}\right)  \, ,\label{e8a}\\
\frac{dq_n}{dt} &=   I_c\sin\varphi_n
+C_J\frac{\hbar}{2e}\frac{d^2\varphi_n}{d t^2}\,.\label{e8b}
\end{alignat}
\end{subequations}
In this case, to the expansion (\ref{re2}) we should add the similar expansion for $q_n$
\begin{eqnarray}
\label{exp}
q_n=A\sin\theta+B\sin 3\theta +D\sin 5\theta+\dots\,.
\end{eqnarray}
Substituting  (\ref{re2}) and (\ref{exp}) into (\ref{e8a}) we obtain
\begin{subequations}
\label{ave88}
\begin{alignat}{4}
\frac{\hbar\omega}{2e} a&=\frac{4}{C}\sin^2\left(\frac{k}{2}\right)A  \, ,\label{ave88a}\\
\frac{3\hbar\omega}{2e} b&=\frac{4}{C}\sin^2\left(\frac{3k}{2}\right)B  \, ,\label{ave88b}\\
\dots  \nonumber
\end{alignat}
\end{subequations}
Substituting  (\ref{re2}) and (\ref{exp}) into (\ref{e8b}) we obtain
\begin{subequations}
\label{ave99}
\begin{alignat}{4}
\omega A &=   I_c\left(a-\frac{a^3}{8}+\frac{a^5}{192}-\frac{a^2b}{8}+\dots\right)
-\frac{\hbar C_J}{2e}\omega^2a\, ,\label{ave99a}\\
3\omega B &= I_c\left(-\frac{a^3}{24}+b+\frac{a^5}{384}-\frac{a^2b}{4}+\dots\right)
-\frac{9\hbar C_J}{2e}b\, ,\label{ave99b}\\
\dots \nonumber
\end{alignat}
\end{subequations}
Combining (\ref{ave88}) and (\ref{ave99}) we can obtain the generalization of Eq. (\ref{zz})

Still another way to obtained the nonlinear dispersion law is based on the averaged Lagrangian \cite{whitham}. The lagrangian of the discrete JTL is \cite{kogan}
\begin{eqnarray}
\label{lag}
L=\frac{C_J\hbar^2}{8e^2}\sum_n\left(\frac{d\varphi_n}{dt}\right)^2
+\frac{\hbar}{2e}\sum_n\frac{dq_n}{dt}\varphi_n  \nonumber\\
-\frac{1}{2C}\sum_n \left(q_{n}-q_{n+1}\right)^2
+\frac{\hbar}{2e}I_c\sum_n\cos\varphi_n \,.
\end{eqnarray}
We substitute into (\ref{lag}) the expansions (\ref{re2}) and (\ref{exp}) and average with respect to $\theta$
\begin{eqnarray}
{\cal L}=\frac{1}{2\pi}\int_0^{2\pi}L(\theta)d\theta\,,
\end{eqnarray}
to obtain
\begin{eqnarray}
\label{begin}
{\cal L}(a,b,d,\dots,A,B,D,\dots)=\frac{C_J\hbar^2}{16e^2}(a^2+9b^2+\dots)\nonumber\\
+\frac{\hbar\omega}{4e}\left(aA+3 bB+\dots\right)\nonumber\\
-\frac{1}{C}\left[\sin^2\left(\frac{k}{2}\right)A^2
+\sin^2\left(\frac{3k}{2}\right)B^2+\dots\right]\nonumber\\
+\frac{\hbar}{8e}I_c \left(-a^2+\frac{a^4}{16}
-\frac{a^6}{576}+\frac{a^3b}{12}\right.\nonumber\\
\left.-b^2-\frac{a^5b}{192}+\frac{a^2b^2}{4}+\dots\right)\,.
\end{eqnarray}
The averaged Lagrangian equations are \cite{whitham}
\begin{subequations}
\label{end}
\begin{alignat}{4}
\frac{\partial {\cal L}}{\partial A}&=0  \, ,\\
\frac{\partial {\cal L}}{\partial B}&=0  \, ,\\
\dots  \nonumber
\end{alignat}
\end{subequations}
and
\begin{subequations}
\label{end2}
\begin{alignat}{4}
\frac{\partial {\cal L}}{\partial a}&=0  \, ,\\
\frac{\partial {\cal L}}{\partial b}&=0  \, ,\\
\dots  \nonumber
\end{alignat}
\end{subequations}
Substituting (\ref{begin}) into (\ref{end}) we  recover (\ref{ave88}), and substituting into (\ref{end2}) - (\ref{ave99}).

\section{Modulation stability}
\label{modul}

The obtained nonlinear dispersion law allows us to study modulation stability of a plane wave.
The slow-envelope wave we can  describe, following  Whitham \cite{whitham},
by the equations
\begin{subequations}
\label{14.14}
\begin{alignat}{4}
\frac{\partial k}{\partial t}+\frac{\partial \omega(k;a^2)}{\partial z}&=0  \, ,\label{14.14a}\\
\frac{\partial a^2}{\partial t}+\frac{\partial \left[\omega'(k;a^2)a^2\right]}{\partial z}&=0 \, , \label{14.14b}
\end{alignat}
\end{subequations}
where $a$ is the amplitude of the envelope, and $k$ is the $z$-derivative of its phase.

Consider modulation of the plane wave with the wavevector $k_0$ and the amplitude $a_0$
\begin{subequations}
\label{14.22}
\begin{alignat}{4}
k&=k_0+k_1(z,t)  \, ,\label{14.22a}\\
a&=a_0+a_1(z,t) \, . \label{14.22b}
\end{alignat}
\end{subequations}
In a frame of reference moving at the group
velocity $\omega'(k_0;a_0^2)$ and after  linearizition with respect to $k_1$ and $a_1$, Eq. (\ref{14.14})   becomes
\begin{subequations}
\label{hr}
\begin{alignat}{4}
\frac{\partial k_1}{\partial t}
+2\omega_2(k_0)a_0\frac{\partial a_1}{\partial z}&=0\, , \label{hra}\\
2a_0\frac{\partial a_1}{\partial t}
+\omega''(k_0;a_0^2)\frac{\partial k_1}{\partial z}a_0^2&=0 \, .\label{hrb}
\end{alignat}
\end{subequations}
We now assume that the perturbations have the form of sinusoidal modulations
with wavenumber $K$ and frequency $\Omega$:
\begin{subequations}
\label{14.23}
\begin{alignat}{4}
k_1&=A\exp\left\{i\left[Kz-\Omega t\right]\right\} + c.c.  \, ,\label{14.23a}\\
a_1&=B\exp\left\{i\left[Kz-\Omega t\right]\right\} + c.c.  \, . \label{14.23b}
\end{alignat}
\end{subequations}
Substituting relations (\ref{14.23}) into (\ref{hr})  we obtain a set of two homogeneous equations
\begin{subequations}
\label{14.25}
\begin{alignat}{4}
\Omega A -2\omega_2(k_0)a_0KB=0  \, ,\label{14.25a}\\
2\Omega B-\omega''(k_0;a_0^2)a_0KA=0 \, . \label{14.25b}
\end{alignat}
\end{subequations}
Equation (\ref{14.25}) has nontrivial solution provided
\begin{eqnarray}
\label{ome}
\Omega^2(K)=\omega_2(k_0)\omega''(k_0;a_0^2)a_0^2K^2\,.
\end{eqnarray}
Notice that (\ref{ome}) is valid for any value of $k_0$, but $K$ is limited by the condition $K\ll 1$.
In our case  $\omega_2(k_0)\omega''(k_0;a_0^2)>0$ (see Eq. (\ref{disper22})), and the plane wave is stable. If the opposite equality  $\omega_2\omega''<0$ were correct,
small perturbations would have grown in time,
and in this sense the plane wave would have been unstable.

We were not able to find in \cite{whitham} the full scale derivation of (\ref{14.14}), nor  were we able to produce it, so we decided at least to compare (\ref{14.14b}) with the equation following from  the nonlinear Schrodinger equation (NLS), which can be written as  \cite{solitons}
\begin{eqnarray}
\label{shr}
i\frac{\partial \psi}{\partial t}+i\omega'(k;|\psi|^2)\frac{\partial \psi}{\partial z}
+\frac{1}{2}\omega''(k;|\psi|^2)\frac{\partial^2 \psi}{\partial z^2}\nonumber\\
-\omega_2(k_0)|\psi|^2\psi=0\,.
\end{eqnarray}
From (\ref{shr}) follows equation for $|\psi|^2$
\begin{eqnarray}
\label{k}
\frac{\partial |\psi|^2}{\partial t}+\omega'(k;|\psi|^2)\frac{\partial |\psi|^2}{\partial z}
+\omega''(k;|\psi|^2)\frac{\partial j}{\partial z}=0\,,
\end{eqnarray}
where
\begin{eqnarray}
j=-\frac{i}{2}\left(\psi^*\frac{\partial \psi}{\partial z}-\psi\frac{\partial \psi^*}{\partial z}\right)\,.
\end{eqnarray}
If we put
\begin{eqnarray}
\label{correct}
\psi=ae^{i\theta} \, ,
\end{eqnarray}
then
\begin{eqnarray}
j=a^2\frac{\partial \theta}{\partial z}\,.
\end{eqnarray}
With this, Eq. (\ref{k}) becomes very close to  (\ref{14.14b}).

\section{Discussion}
\label{discussion}

Recently,  quantum mechanical description of JTL in general and parametric amplification in such lines in particular started to be developed, based on
quantisation techniques in terms
of discrete mode operators \cite{reep},  continuous  mode operators \cite{fasolo},
 a Hamiltonian approach  in the Heisenberg and interaction pictures \cite{greco},
 the quantum Langevin method \cite{yuan}, or on partitions a quantum device into compact lumped or quasi-distributed cells \cite{minev}.
It  would be interesting to understand in what way the results of the present paper are changed by quantum mechanics.
Particularly interesting looks studying  of
quantum ripples over a semi-classical shock \cite{glazman} and fate of quantum shock waves at late times \cite{glazman2}.
Closely connected problem of classical and quantum dispersion-free coherent propagation
in waveguides and optical fibers was studied recently in Ref. \cite{costas}.
Also, it would be interesting to study how the results obtained in the paper change, when the current phase relation is generalized \cite{zutic}.

Finally, we would like to express our
hope that the results obtained in the paper are  applicable to kinetic inductance based traveling wave parametric amplifiers based on a coplanar waveguide architecture.
Onset of shock-waves in
such amplifiers is an undesirable phenomenon. Therefore, shock
waves in various JTL should be further studied, which was one of motivations of the present work.

\begin{acknowledgments}

The main idea of the present work was born in the discussions with M. Goldstein.
We are also grateful to  J. Cuevas-Maraver, A. Dikande, M. Inc, P. Kevrekidis, B. A. Malomed,  V. Nesterenko, T. H. A. van der Reep,  B. Ya. Shapiro, A.  Vainchtein, and I. Zutic for their comments (some of which were  crucial for the
completion of the project) and to P. Rosenau for his criticism.

\end{acknowledgments}

\begin{appendix}

\section{Propagator for the linear transmission line}
\label{linear}

In this Section we consider the transmission line, obtained from that presented on Fig. \ref{trans1}, by substituting
linear inductor for the JJ.
The  circuit equations are
\begin{subequations}
\label{l}
\begin{alignat}{4}
L\frac{d I_n}{d t}&=\frac{1}{C}\left(q_{n+1}-2q_{n}+q_{n-1}\right)  \, ,\label{la}\\
\frac{dq_n}{dt} &=   I_n \, ,\label{lb}
\end{alignat}
\end{subequations}
where  $I_n$ is the current,  $C$ is the capacitance, and  $L$ is the inductance.
Eliminating $I_n$ and introducing the dimensionless time $\tau=t/\sqrt{LC}$ we obtain
\begin{eqnarray}
\label{22}
\frac{d^2q_n(\tau)}{d\tau^2} =q_{n+1}(\tau)-2q_n(\tau)+q_{n-1}(\tau)\, .
\end{eqnarray}
Because the system is linear (but dispersive), it doesn't
allow either kinks or solitary waves, and thus seems to
lie outside the scope of the paper. However, we'll use
the system to  check up the modified quasi-continuum approximation, which  Section~\ref{contr} we apply to the JTL.

\subsection{The exact solution}

We  define the propagator by the  initial and the boundary conditions
\begin{subequations}
\label{condi1}
\begin{alignat}{4}
q_n(0)=\delta_{n0}  \, , \hskip 1cm  \dot{q}_n&(0)=0\,,\\
\lim_{n\to\pm\infty}  q_n=&0\,.
\end{alignat}
\end{subequations}

Recalling the recurrence relation satisfied by Bessel functions \cite{abram}
\begin{eqnarray}
2\frac{dZ_n(\tau)}{d\tau}=Z_{n-1}(\tau)-Z_{n+1}(\tau)  \,,
\end{eqnarray}
where $Z$ is any Bessel function,  and repeating it twice we obtain
\begin{eqnarray}
\label{plau}
4\frac{d^2Z_n(\tau)}{d\tau^2}=Z_{n+2}(\tau)-2Z_{n}(\tau)+Z_{n-2}(\tau)  \,.
\end{eqnarray}
Comparing (\ref{plau}) with (\ref{22}) we obtain  plausible solution
for half of the problem.  This solution -- for even $n$ --
is
\begin{eqnarray}
\label{b2}
q_n(\tau)=J_{2n}(2\tau)\,,
\end{eqnarray}
where $J_n$ is the Bessel function of the first kind.

To obtain a rigorous solution (and for the whole problem) we
use  Laplace transformation
\begin{equation}
Q_n(s) =  \int_0^\infty {\rm d} \tau\, e^{-s\tau} q_n(\tau) \, .
\end{equation}
For $Q_n(s)$ we obtain the difference equation
\begin{equation}
\label{dif}
Q_{n+1}(s)-(2+s^2) Q_n(s)+ Q_{n-1}(s)=-s\delta_{n0}\,.
\end{equation}
Solving (\ref{dif}) we get
\begin{equation}
\label{bro}
Q_{n}(s)= \frac{1}{\sqrt{s^2+4}}\left(\frac{\sqrt{s^2+4}-s}{2}\right)^{2|n|}\,.
\end{equation}
Taking into account  the  inverse Laplace transform
 correspondence tables \cite{abram},
we obtain Eq. (\ref{b2}) for all $n$.

Though we will not use the following result, consider the signalling in
the discrete semi-infinite linear transmission line. The problem
is characterized by Eq. (\ref{22}) for $n\geq 1$ with
the   initial
and the boundary conditions
\begin{subequations}
\label{condi2}
\begin{alignat}{4}
&q_n(0)=\dot{q}_n(0)=0\,,\\
q_0(\tau)=&\delta(\tau)\,,\hskip .7cm \lim_{n\to+\infty}  q_n(\tau)=0\,.
\end{alignat}
\end{subequations}

The problem can be solved exactly.
After Laplace transformation
 we obtain  difference equation
\begin{equation}
\label{dif2}
Q_{n+1}(s)-(2+s^2)Q_n(s)+Q_{n-1}(s)=0
\end{equation}
with the boundary conditions
\begin{eqnarray}
Q_0(s)=1\,,\hskip 1cm \lim_{n\to+\infty}  Q_n(\tau)=0\,.
\end{eqnarray}
Solving (\ref{dif2}) we get
\begin{equation}
\label{broa}
Q_{n}(s)= \left(\frac{\sqrt{s^2+4}-s}{2}\right)^{2n} \, .
\end{equation}
Taking into account  the  inverse Laplace transform
 correspondence tables \cite{abram},
we obtain \cite{bulla,kogan}
\begin{eqnarray}
q_n(\tau)=\frac{2n}{\tau}J_{2n}(2\tau)\,.
\end{eqnarray}

\subsection{The modified quasi-continuum approximation}
\label{trun}

Now let us  solve the problem approximately. We'll consider $q$ as a function of the continuous variable $z=n$ (for simplicity in this Section and in the next one we put $\Lambda=1$), and  present the r.h.a. of Eq. (\ref{22}) modifying the quasi-continuum approximation (\ref{comm33b}) to
\begin{eqnarray}
\label{commb2}
q_{n+1}(\tau)-2q_n(\tau)+q_{n-1}(\tau)=\left(\frac{\partial}{\partial z}+\frac{1}{24}\frac{\partial^3}{\partial z^3}\right)^2q\,.\nonumber\\
\end{eqnarray}
We will call (\ref{commb2}) the modified quasi-continuum approximation.
After that, (\ref{22}) is decoupled  into two equations for right and left going waves
\begin{eqnarray}
\label{ve8}
\frac{\partial q}{\partial \tau}=\pm
\left(\frac{\partial q}{\partial z}+\frac{1}{24}\frac{\partial^3 q}{\partial z^3}\right)\,.
\end{eqnarray}

The propagator is defined by the initial
and the boundary conditions
\begin{eqnarray}
q(z,0)=\delta(z)  \, , \hskip 1cm
\lim_{z\to\pm\infty} q(z,\tau)=0\,.
\end{eqnarray}
Making Laplace transformation with respect to $\tau$ and Fourier transformation with respect to $z$
\begin{eqnarray}
{\cal Q}(k,s)=\int_0^{\infty}d\tau e^{-s/\tau}\int_{-\infty}^{+\infty}dz q(z,\tau)e^{ikz}\,,
\end{eqnarray}
we obtain for the right going part of the propagator the equation
\begin{eqnarray}
\label{c6}
\left(s -ik+\frac{ik^3}{24}\right){\cal Q}(k,s)=1\,.
\end{eqnarray}
Solving Eq. (\ref{c6}) we get
\begin{eqnarray}
\label{c7}
{\cal Q}(k,s)=\frac{1}{s -ik+\frac{ik^3}{24}}\,.
\end{eqnarray}
Making  the  inverse Laplace and Fourier transformations
we obtain
\begin{eqnarray}
\label{d95}
q(z,\tau)&=&\frac{1}{4\pi}
\int_{-\infty}^{+\infty}dk \exp[i(\tau-z)k-i\tau k^3/24]
\nonumber\\
&=&\tau^{-1/3}\text{Ai}\left[2\tau^{-1/3}(z-\tau)\right]\,,
\end{eqnarray}
where Ai is the   Airy function \cite{abram}.
 Equation (\ref{d95})
describes  the signal front at $z\sim\tau/2$,
exponentially small precursor for
$\tau<2z$, and oscillations and power law decrease of the
signal  in the wake for  $\tau>2z$. The width of the
transition region between the two asymptotic forms increases
with time as $\tau^{1/3}$.

\begin{figure}[h]
\includegraphics[width=.8\columnwidth]{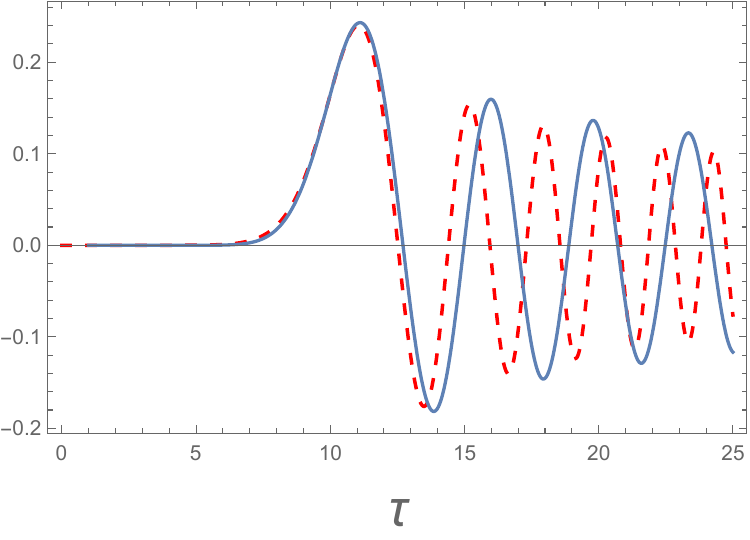}
\caption{Propagator calculated for $n=10$  exactly (Eq. (\ref{b2}), solid blue line) and for $z=10$ in the framework of the  modified quasi-continuum approximation  (Eq. (\ref{d95}), dashed red line). }
 \label{trans13}
\end{figure}

Fig. \ref{trans13}  compares   Eq. (\ref{d95}) with the exact result (\ref{b2})
for $\tau$ from zero up to a couple of $z$. To compare the results for $\tau\gg z$, we may use asymptotic forms of Bessel and Airy functions \cite{abram}
\begin{subequations}
\begin{alignat}{4}
&J_{2n}(2\tau)\sim \sqrt{\frac{1}{\pi\tau}}(-1)^n\cos\left(2\tau-\frac{\pi}{4}\right)\,,\\
&\tau^{-1/3}\text{Ai}\left[2\tau^{-1/3}(z-\tau)\right]\sim \sqrt{\frac{1}{\pi\tau}}\cos\left[A\tau
-\frac{\pi}{4}\right]\,,
\end{alignat}
\end{subequations}
where
$A=2^{5/2}/3\approx 1.9$.

\section{The integral approximation: the kinks}
\label{int}

In this Appendix we are looking for some  way to approximate the finite difference in the r.h.s. of Eq.  (\ref{ave7a}) alternative to Taylor expansion (\ref{comm33}).
We were not able to advance far on the road we have taken here (if at all). However, some equations obtained in the process look quite amusing to us, and we decided to present them to general attention.

Treating $\varphi$ and $q$ as  functions of the continuous variable $z$ (which we measure in $\Lambda$), let us approximate the finite difference in the r.h.s. of Eq.  (\ref{ave7a}) as
\begin{eqnarray}
\label{omm4}
q_{n+1}-2q_{n}+q_{n-1} =
\int_{-\infty}^{+\infty} dz'g(z-z')\frac{d^2q(z',\tau)}{dz'^2}\,,
\end{eqnarray}
where $g(z)$ is a  non-singular function, which
is positive, even and
has the following zero and second moments
\begin{subequations}
\begin{alignat}{4}
\int_{-\infty}^{+\infty} dzg(z)&=1  \, ,\label{zero}\\
\int_{-\infty}^{+\infty} dzz^2g(z)&=\frac{\Lambda^2}{6}\, ,
\end{alignat}
\end{subequations}

Looking
for the running wave (\ref{run}) solution of (\ref{ave7}),
we obtain the integro-differential equation for the function $\varphi(x)$
\begin{eqnarray}
\label{v78}
\overline{U}^2\frac{d\varphi(x)}{d x}=\int_{-\infty}^{+\infty} dx'\frac{dg(x-x')}{dx}\sin\varphi(x')\,.
\end{eqnarray}
Integrating Eq. (\ref{v78}) with respect to $x$
we obtain  nonlinear Fredholm integral equations of the second
kind \cite{wazwaz}
\begin{eqnarray}
\label{v79}
\overline{U}^2\varphi(x)=\int_{-\infty}^{+\infty} dx'g(x-x')\sin\varphi(x')-F\,.
\end{eqnarray}
Imposing  the boundary conditions (\ref{gran}) and going to the limits
$x\to +\infty$ and $x\to -\infty$,
we recover Eq. (\ref{v797})  and, hence, (\ref{velocity}) and (\ref{vb}).
Substituting $\overline{U}^2$ and $F$ into Eq. (\ref{v79}) we get the counterpart of  Eq. (\ref{v9}) (or (\ref{v99}))
\begin{eqnarray}
\label{v74}
\varphi(x)&=&\frac{\varphi_1-\varphi_2}{\sin\varphi_1-\sin\varphi_2}
\int_{-\infty}^{+\infty} dx'g(x-x')\sin\varphi(x')\nonumber\\
&+&\frac{\varphi_2\sin\varphi_1-\varphi_1\sin\varphi_2}{\sin\varphi_1-\sin\varphi_2}\,.
\end{eqnarray}

Now let us consider Eq. (\ref{v74}) per se, forgetting the properties of $\varphi(x)$ which were postulated to  derive it. We realise that
if $\varphi(x)$  goes to some limits when $x\to +\infty$ and $x\to -\infty$, each of these limits is either $\varphi_1$, or $\varphi_2$. This is unfortunately all we can say about the solution. Previously we have seen that Eq. (\ref{v9}) (or (\ref{v99})) has solution only if $\varphi_2=-\varphi_1$. We are unable to prove that for Eq. (\ref{v74}).
However, if the relation $\varphi_2=-\varphi_1$ is imposed,
Eq. (\ref{v74}) takes the form
\begin{eqnarray}
\label{v79n}
\varphi(x)=\frac{\varphi_1}{\sin\varphi_1}\int_{-\infty}^{+\infty} dx'g(x-x')\sin\varphi(x')\,.
\end{eqnarray}
The only thing we can prove about the solution of Eq. (\ref{v79n}) is that, for any $x$,
\begin{eqnarray}
\label{equ}
-\varphi_1\leq \varphi(x)\leq \varphi_1
\end{eqnarray}
(for the sake of definiteness we consider $\varphi_1$ to be positive). In fact, let
$\sin\varphi(x)$ reaches maximum value
at some point $x_0$, and  $\sin\varphi(x_0)>\sin\varphi_1$. Then
\begin{eqnarray}
\label{vn}
\frac{\varphi_1}{\sin\varphi_1}\int_{-\infty}^{+\infty} dx'g(x_0-x')\sin\varphi(x')
\nonumber\\
<\frac{\varphi_1}{\sin\varphi_1}\sin\varphi(x_0)<\varphi_0
\end{eqnarray}
(in the last step we took into account that $\sin\varphi/\varphi$ decreases when $\sin\varphi$ increases for positive $\varphi$). So we came to a contradiction. Similar for the minimum value of $\sin\varphi$.

\section{A couple of additions (written after the paper was published)}
\label{ape}

We would like also to use the opportunity and to add that Eq. (\ref{ph}) is more general than the assumptions used to derive it in the body of the paper.
Let us return to Eq. (\ref{v90}).
Multiplying both sides by $d\sin\varphi/dx$
we may   integrate it
to obtain
\begin{eqnarray}
\label{v10d}
\int\left[\frac{\Lambda^2}{12}\frac{d^2\sin\varphi}{d x^2}\dots\right]\frac{d\sin\varphi}{dx}dx
=-\Pi(\sin\varphi)+E\,,
\end{eqnarray}
where $\Pi(\sin\varphi)$ is given by Eq.
(\ref{v10b}).
The l.h.s. of  (\ref{v10d})
can be obtained on the basis of relations
\begin{subequations}
\begin{alignat}{4}
\frac{d^2y}{d x^2}\frac{dy}{d x}&=\frac{1}{2}\frac{d}{dx}\left(\frac{dy}{d x}\right)^2  \, ,\\
\frac{d^4y}{d x^4}\frac{dy}{d x} &=\frac{d}{dx}\left[\frac{d^3y}{d x^3}\frac{dy}{d x}-\frac{1}{2}\left(\frac{d^2y}{d x^2}\right)^2\right] \, ,\\
\frac{d^6y}{d x^6}\frac{dy}{d x} &=\frac{d}{dx}\left[\frac{d^5y}{d x^3}\frac{dy}{d x}-\frac{d^4y}{d x^4}\frac{d^2y}{d x^2}+\frac{1}{2}\left(\frac{d^3y}{d x^3}\right)^2\right] \, ,
\end{alignat}
\end{subequations}
and, in general,
\begin{eqnarray}
\label{r}
\frac{d^{2m}y}{d x^{2m}}\frac{dy}{d x}=\frac{d}{dx}\left[\frac{d^{2m-1}y}{d x^{2m-1}}\frac{dy}{d x}-\frac{d^{2m-2}y}{d x^{2m-2}}\frac{d^2y}{d x^2}
\right.\nonumber\\
+\left.\frac{d^{2m-3}y}{d x^{2m-3}}\frac{d^3y}{d x^3}-\dots-\frac{1}{2}(-1)^m\left(\frac{d^my}{d x^m}\right)^2\right]\,.
\end{eqnarray}

The solutions, we are interested in, are characterised by the boundary conditions
(\ref{gran}).
From the structure of Eq. (\ref{r}) follows that for the asymptotic values of $x$, the  l.h.s. of Eq. (\ref{v10d}) is equal to zero. Thus we regain (\ref{phii}),  which
in the main body of the paper we derived   after truncating the series (\ref{comm33}) to (\ref{comm33b}). Now we see that the truncation was not necessary.
From (\ref{phii}) and (\ref{v797}) we recover Eq. (\ref{ph}).

And now the final addition. Let us return to Eq. (\ref{v10}), in which, for the case of the soliton, $\Pi(\sin(\varphi))$ is given by Eq. (\ref{ss2}) and $E=0$.
In the main body of the paper we considered  solitons, weak in the sense $|\varphi_1|\sim 1$,
$|\varphi_1-\varphi_0|\ll 1$. In this Appendix we would like to consider solitons weak in the sense $|\varphi_1|,|\varphi_0|\ll 1$, and, hence, also $|\varphi|\ll 1$. In this case Eq. (\ref{v10}) takes the form
\begin{eqnarray}
\label{v10e}
\Lambda^2\left(\frac{d \varphi}{d x}\right)^2=(\varphi_1-\varphi)^2(\varphi-\varphi_0)
(\varphi_0+2\varphi_1+\varphi)\,.
\end{eqnarray}

\end{appendix}

\end{document}